\documentclass[twocolumn,showpacs,preprintnumbers,amsmath,amssymb,final]{revtex4}

\usepackage{graphicx}
\usepackage{dcolumn}
\usepackage{bm}
\usepackage[latin1]{inputenc}
\usepackage{epsfig}

\begin{document}

\title{ Resonant invisibility with finite range interacting fermions}
\author{Jean-Pierre Nguenang$^{1,2}$}
\author{Sergej Flach$^{1}$}
\author{Ramaz Khomeriki$^{1,3}$}
\affiliation{$^1$Max-Planck-Institut f\"ur Physik komplexer
Systeme, N\"othnitzer Str. 38, 01187 Dresden, Germany}
\affiliation{$^2$Fundamental physics laboratory: Group of
Nonlinear physics and Complex systems, Department of Physics,
University of Douala, P.O. Box 24157, Douala-Cameroon }
\affiliation{$^3$Department of Physics, Tbilisi State University,
3 Chavchavadze, 0128 Tbilisi, Georgia }

\date{\today}

\begin{abstract}
We study the eigenstates of two opposite spin fermions on a one-dimensional lattice
with finite range interaction.
The eigenstates are projected onto the set of Fock eigenstates of the noninteracting case.
We find antiresonances for symmetric eigenstates, which eliminate the interaction between
two symmetric Fock states when satisfying a corresponding selection rule.

\end{abstract}


\maketitle

{\it Introduction.} Up to now a lot of interest in experiments on
cold atoms has been focused on matter wave properties of the
condensates which are described by the Hartree-Fock-Bogoliubov
mean field model for weakly interacting quantum gases
\cite{Nature416,Dalfogo71,Legett73,Pitaevskii2003,Pethick2001}. At
the same time, the use of collision processes turns out to be a promising
approach to implement quantum gate operations \cite{Mendel2003}. A
standard method for the description of such systems is to map them
to Hubbard like lattice models where atomic physics provides a
whole toolbox to engineer various types of Hamiltonians for 1D,
2D, and 3D Bose and Fermi systems.

The interplay of interactions and discreteness leads to a set of
interesting phenomena, including bound states, see e.g.
\cite{Fleurov,Scott1,Pinto,Pinto2007,Pinto2008} and
\cite{EilbeckPhysicaD78,Dorignac2004,Pouthier2003PRE}. In recent
papers \cite{NguenangPRB75,jpnflach09PhysRevA80} we have studied
properties of such bound states (also frequently coined quantum
breathers) in various one dimensional Hubbard like models by
considering two bosons or two fermions (with opposite spins) on
lattices. The fermionic case adds to the complexity with the spin
as an additional degree of freedom. Consequently two fermions can
form up to three different bound states, while two bosons form
only one. In all these cases the interaction was assumed to be
local, i.e. both particles interact only when occupying the same
lattice site. In the present paper we consider fermionic particles
with a finite range of interaction, as a more realistic
description of experimental situations, which may be directly
applicable in quantum computing, where the controlled interaction
can be used to create entanglement with high fidelity. We analyze
two particle eigenfunctions and identify resonance conditions for
which two particles do not scatter despite the presence of a
nonzero interaction.

{\it Model and spectrum.} We consider one-dimensional periodic
lattice with $f$ sites described by an extended  fermionic
Bose-Hubbard (EFBH) model. The Hamiltonian is given by
\begin{eqnarray}\label{eq:hamiltonian}
\hat{H} = \hat{H}_0 +  \hat{H}_U + \hat{H}_V, \\
\hat{H}_{0}=-\sum_{j,\sigma}\hat{a}_{j,\sigma}^+(\hat{a}_{j-1,\sigma}
+\hat{a}_{j+1,\sigma}),  \\
 \hat{H}_U = -U\sum_{j}
\hat{n}_{j,\uparrow}\hat{n}_{j,\downarrow},{\ } {\ }{\ }{\ }{\
}\hat{n}_{j,\sigma}= \hat{a}_{j,\sigma}^+\hat{a}_{j,\sigma},
\label{eq:haminteraction1}
\end{eqnarray}
\begin{equation}
\hat{H}_V = -V\sum_{j}\hat{n}_j\hat{n}_{j+1},  {\ }{\ }{\ }
\hat{n}_j=\hat{n}_{j,\uparrow}+\hat{n}_{j,\downarrow}.
\label{eq:haminteraction2}
\end{equation}
Here $\hat{H}_0$ describes the nearest-neighbor hopping,
$\sigma=\uparrow$,$\downarrow$ denotes the spin, $\hat{H}_U$ and
$\hat{H}_V$ describe the onsite and intersite (between adjacent
sites) interaction between the particles with strengths $U$, and
$V$, respectively; $a_{j,{\sigma}}^+$ and $a_{j,{\sigma}}$ are the
fermionic creation and annihilation operators  satisfying the
anticommutation relations:
$[\hat{a}_{j,{\sigma}},\hat{a}_{l,{\sigma}'}^+]=\delta_{j,l}\delta_{\sigma,\sigma'}$,
and
$[\hat{a}_{j,{\sigma}},\hat{a}_{l,\sigma'}]=[\hat{a}_{j,{\sigma}}^+,\hat{a}_{l,\sigma'}^+]=0$.
The sign of $U$ and $V$ is not specified. The Hamiltonian
(\ref{eq:hamiltonian}) commutes with the number operator
 $\hat{N}=\sum_{j}\hat{n}_j$  whose
eigenvalues are $n=n_{\uparrow}+n_{\downarrow}$, which is the total number of fermions in
the lattice. In this work we focus on the simplest  nontrivial case of $n=2$,
with $n_{\uparrow}=1$ and $n_{\downarrow}=1$.

To describe quantum states, we use a number state basis
$|\Phi_n\rangle=|n_1;n_2\cdots n_f\rangle$
\cite{EilbeckPhysicaD78}, where $n_i=n_{i,{\uparrow}}+
n_{i,{\downarrow}}$ represents the number of fermions  at the i-th
site of the lattice. The fermionic two particle states are
generated from the vacuum $|O\rangle$ by successively creating a
particle with spin down and spin up.

The Hamiltonian (\ref{eq:hamiltonian}) commutes with the
translational operator $\hat{T}$, which shifts all lattice indices
by one. Its eigenvalues are $\tau=exp(ik)$ with the Bloch wave
number $k=\frac{2\pi\nu}{f}$, and $\nu=0,1,2,\cdots,f-1$ .

\noindent {\it Single-fermion states.} In this simplest case,
only one fermion is in the lattice (either with spin up or down)
($n=1$), and the state is represented by
$|j\rangle=\hat{a}_{j,\sigma}^+|O\rangle$. The interaction terms
($\hat{H}_U$ and $\hat{H}_V$ ) have no contribution for a single
particle. Thus the eigenstates of the Hamiltonian
(\ref{eq:hamiltonian}) are the eigenstates of $\hat{H}_0$ which
are given by:
\begin{equation}
|\Psi_1\rangle=\frac{1}{\sqrt{f}}
\sum_{s=1}^f\Big(\frac{\hat{T}}{\tau}\Big)^{s-1}|1\rangle \;.
\end{equation}
The corresponding eigenenergies are
\begin{equation}\label{eq:1fermionenergy}
\varepsilon_k=-2\cos(k)\;.
\end{equation}

\noindent {\it Two fermions.} For the case of two opposite spin
fermions ($n=2$ with $n_{\uparrow}=1$ and $n_{\downarrow}=1$),
each eigenstate is formed as a linear combination of number states
with fixed $n$.
\begin{equation}
|\Psi_n\rangle=\sum_{j}c_j|\Phi_n^j\rangle\;.
\end{equation}
For two particles,  this involves $N_{s}=
f^2$  basis states, $|\Phi_2^j\rangle $,  which is the number of
ways one can distribute two fermions with opposite spins over the
$f$ sites with possible double occupancy. Then we define the basis
state with a given value of the wave number $k$ as in Ref.
\cite{jpnflach09PhysRevA80} and a complete wave function  is:
\begin{equation}\label{eq:basis1}
|\Psi_2^k\rangle=
c_1|\Phi_1\rangle+\sum_{j=2}^{(f+1)/2}c_{j,+}|\Phi_{j,+}\rangle
              +\sum_{j=2}^{(f+1)/2}c_{j,-}|\Phi_{j},-\rangle\;.
\end{equation}
Any vector in the corresponding Hilbert space is spanned by the
numbers $|c_{1}, c_{2,+},c_{2,-},c_{3,+},c_{3,-}\cdots\rangle$ and
the vectors $|\Phi_{1}\rangle$, $|\Phi_{j,+}\rangle$ and
$|\Phi_{j,-}\rangle$ in two fermion case are defined as follows:
\begin{eqnarray} \label{dop}
|\Phi_1\rangle=\frac{1}{\sqrt{f}}
\sum_{s=1}^f\Big(\frac{\hat{T}}{\tau}\Big)^{s-1}\hat{a}_{1,\uparrow}^+
\hat{a}_{1,\downarrow}^+|O\rangle; \nonumber \\
|\Phi_{j,+}\rangle=\frac{1}{\sqrt{f}}
\sum_{s=1}^f\Big(\frac{\hat{T}}{\tau}\Big)^{s-1}\hat{a}_{j,\uparrow}^+
\hat{a}_{1,\downarrow}^+|O\rangle; \\
|\Phi_{j,-}\rangle=\frac{1}{\sqrt{f}}
\sum_{s=1}^f\Big(\frac{\hat{T}}{\tau}\Big)^{s-1}
\hat{a}_{1,\uparrow}^+\hat{a}_{j,\downarrow}^+|O\rangle; \nonumber
\end{eqnarray}

We diagonalize the Hamiltonian (\ref{eq:hamiltonian}) in the
framework of the  basis defined  in (\ref{eq:basis1}) and derive
the eigenenergies for each given Bloch wave number $k$ from
$\hat{H}|\Psi_2^k\rangle=E|\Psi_2^k\rangle$. This leads to an $f\times f$ matrix
whose elements $H_{i,j}$ ($i,j=2,\ldots,(f+1)/2$) are derived from
\begin{equation}\label{eq:matel}
H_{i,1}=H_{1,i}^*=\langle\Phi_{i,\pm}|\hat{H}|\Phi_{1}\rangle\;,\;
H_{i,j}=\langle\Phi_{i,\pm}|\hat{H}|\Phi_{j,\pm}\rangle\;.
\end{equation}

\begin{figure}[t]
\includegraphics[width=3.2in]{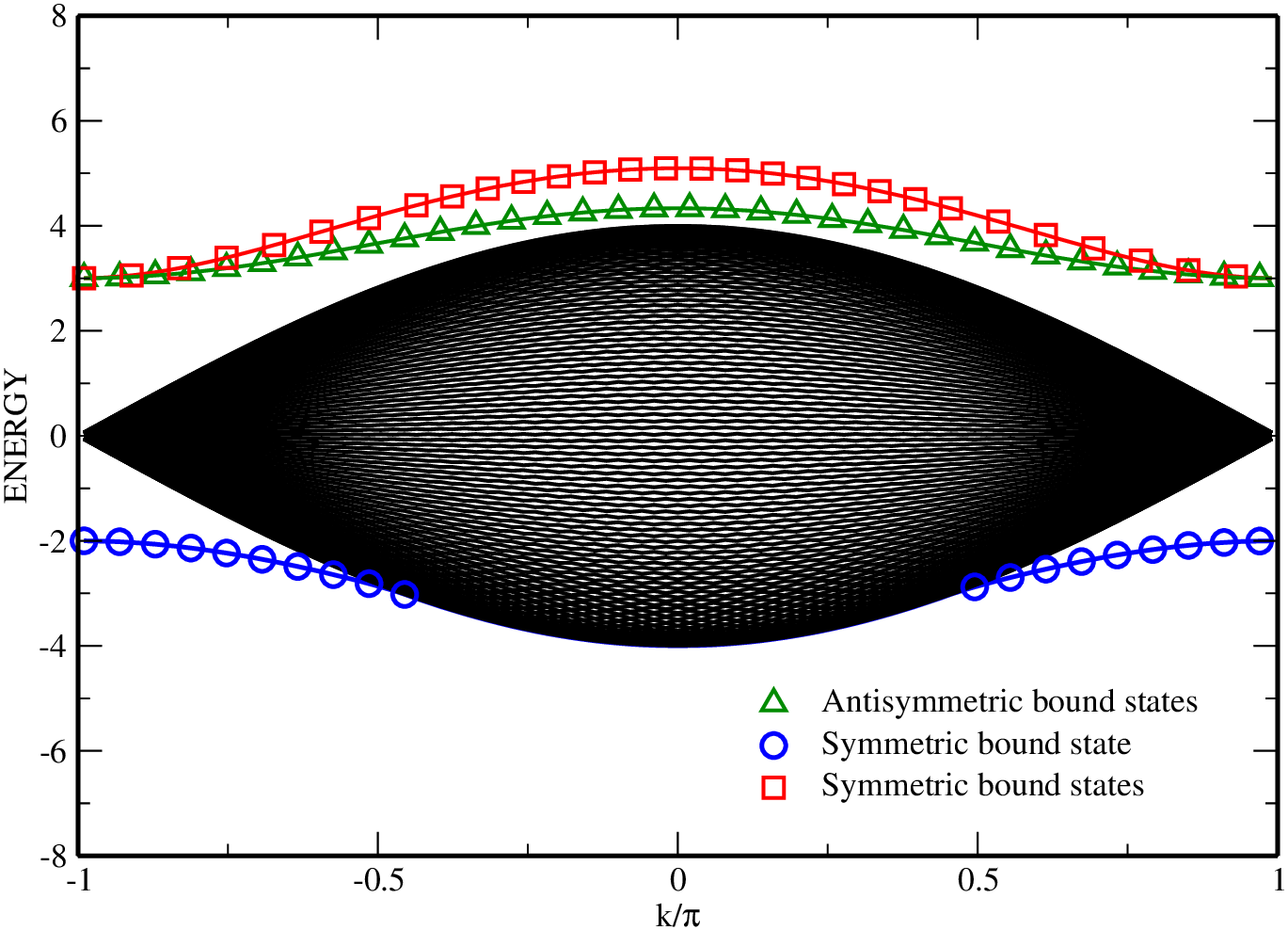}
\caption{\label{spectrum} Energy spectrum of two fermions of the
EFBH chain with periodic boundary conditions for $U=2$, $V=-3$ and
$f=101$. The lines follow from numerical diagonalization of the
matrix \ref{eq:matel} and symbols are the results of analytical
computations for the bound states similar to the calculations in
\cite{jpnflach09PhysRevA80}.}
\end{figure}

We show in Fig. \ref{spectrum} the energy spectrum of the
Hamiltonian matrix (\ref{eq:matel}) obtained by numerical
diagonalization for the case of opposite signs of interaction
parameters $U=2$ and $V=-3$ and the form of the spectrum is
similar to the one in Ref. \cite{jpnflach09PhysRevA80}. Besides a
two particle continuum, three bound state bands are found. The
eigenstates $|{\Phi}_{k_1,k_2}\rangle$ of the continuum correspond
to two fermions independently moving along the lattice as with
zero interaction, and are derived from (\ref{eq:basis1}). Their
eigenenergies are given by :
\begin{equation}\label{eq:energyunperturbed}
E_{k,k_1}^0=-4\cos(k/2)\cdot\cos(k_1),
\end{equation}
with $ {k}$ being the Bloch wave number and
$k_{1}={2\pi}\nu/(f-1)$ , being the canonically conjugated momentum of the
relative coordinate (distance) between  both particles and $\nu=
0,\ldots,(f-1)/2$. Equation (\ref{eq:energyunperturbed}) is the
result of the sum of Bloch bands $E_\pm =-2\cos(\frac{k}{2}\pm
k_1)$ of two asymptotically free particles \cite{MValiente}.

{\it Weight functions in normal mode space.} We
transform to the basis of the symmetric and antisymmetric
states
\begin{equation}\label{eq:basisym}
|\Phi_{j,s}\rangle=\frac{|\Phi_{j,+}\rangle +
|\Phi_{j,-}\rangle}{\sqrt{2}}, \quad
|\Phi_{j,a}\rangle=\frac{|\Phi_{j,+}\rangle -
|\Phi_{j,-}\rangle}{\sqrt{2}}
\end{equation}
where $a$ and $s$ refer to the antisymmetric and the symmetric
states, respectively, $j=2,\ldots,(f+1)/2$. Note that
$|\Phi_1\rangle$ is also a symmetric state. In this basis the
matrix (\ref{eq:matel}) decomposes into two irreducible parts
given by
\begin{eqnarray}\label{eq:matsym}
H^s(i,j) = -\left( \begin{array}{ccccccc}
U & q\sqrt{2} & & & & \\
q^*\sqrt{2} &V & q & & & \\
      & q^* & 0 & q & & \\
  &  & \ddots & \ddots &\ddots & \\
  &  & & q^* & 0 & q \\
  &  & &  & q^* & p
\end{array} \right),
\end{eqnarray}
and
\begin{eqnarray}\label{eq:matasym}
H^a(i,j) = -\left( \begin{array}{cccccc}
V & q & & & & \\
q^* & 0 & q & & & \\
      & q^* & 0 & q & & \\
  &  & \ddots & \ddots &\ddots & \\
  &  & & q^* & 0 & q \\
  &  & &  & q^* &- p
\end{array} \right),
\end{eqnarray}
with $q=1+\tau$ and $p=\tau^{-(f+1)/2} + \tau^{-(f-1)/2}$. The
rank of the symmetric matrix is $(f+1)/2$ and the rank of the
antisymmetric matrix is $(f-1)/2$.

Our strategy is to compute an eigenstate for the interacting case, and
use a weight function to expand it in the basis of the eigenstates of the
noninteracting case.
For this
purpose we fix the Bloch momentum $k$, and choose a seed eigenstate
$|\Psi_{\tilde{k}_1}^0\rangle$ of the unperturbed case with seed mode number
$\tilde{k}$. Upon
switching on the interaction it becomes a new eigenstate
$|\Psi_{\tilde{k}_1}\rangle$, which will have overlap with several
eigenstates of the unperturbed case. We expand the eigenfunction
of the perturbed system using first order perturbation approximation:
\begin{equation} \label{122}
|\Psi_{\tilde{k}_1}\rangle=|\Psi_{\tilde{k}_1}^0\rangle+\epsilon\sum_{{{k}_1^{\prime}}\neq
\tilde{k}_1}\frac{\langle\Psi_{k_1^{\prime}}^{0}|\hat{H}_U+\hat{H}_V|\Psi_{\tilde{k}_1}^0\rangle}
{{E_{\tilde{k}_1}^0}-E_{{k}_1^{\prime}}^0}|\Psi_{{k}_1^{\prime}}^0\rangle.
\end{equation}
From expansion (\ref{122}) it follows that the off-diagonal
($k_1\neq\tilde{k}_1$) weight function at the first order is given
by :
\begin{equation} \label{111}
C^s(k_1;\tilde{k}_1)
=\frac{(2U|q|^2+VE_{k_1}^0E_{\tilde{k}_1}^0)^2|\langle\Psi_{k_1}^0|\Psi_{\tilde{k}_1}^0\rangle|^2}
{4|q|^4(E_{\tilde{k}_1}^0-E_{{k}_1}^0)^2}.
\end{equation}
with $|q |^2=2+2\cos(k)$ and $E_{k_1}^0$ and $E_{\tilde{k}_1}^0$
are the eigenenergies of the unperturbed system given by Eq.
(\ref{eq:energyunperturbed}).
\begin{figure}[top]
\includegraphics[width=2.in]{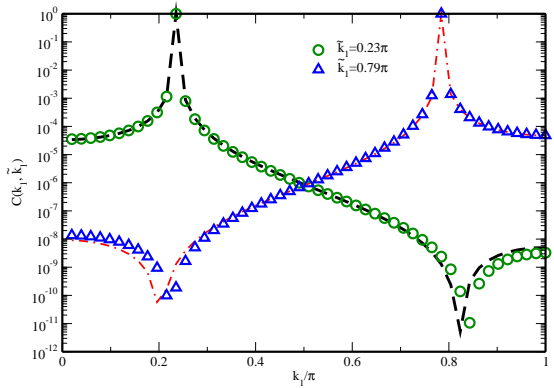}
\caption{\label{gcorr4} Weight function for two seed mode numbers
$\tilde{k}_1=0.23\pi$ and $\tilde{k}_1=0.79\pi$, with the
onsite and intersite interaction parameters $U=0.1$ and $V=0.08$. Here
 $k=0.5\pi$ and $f=101$. Dashed lines are the results using formula (\ref{111})}
\end{figure}

{\it Symmetric
states.}
First we consider the case of small interaction constants
and as expected we find localization in normal mode space. For
instance, in the case of dominant onsite interaction constant
$U=1$ and $V=0.1$, quite similar results to those obtained in Ref.
\cite{NguenangPRB75} are derived and the result of the
perturbation formula (\ref{111}) matches pretty well with those of
the diagonalization procedure.

But if now we take interaction constants $U,V$ with comparable values
(but again in perturbative limit) one will unavoidable deal with
additional antiresonance structure presented in Fig. \ref{gcorr4},
where the weight function vanishes exactly.
The appearance of these structures follows from
the analytical formula (\ref{111}). Indeed, in perturbative limit
of interaction constants one can find such a seed $\tilde k_1$ and
probe $k_1$ wavenumbers that the weight function becomes exactly zero.
We find the following condition for zero weight:
 \begin{equation}\label{eq:resonance}
\frac{U}{V}=-2\cos(k_1)cos(\tilde{k}_1)  \;
\end{equation}
It also follows from eq. (\ref{eq:resonance}) that there is a critical
wave number given by
\begin{equation}\label{eq:criticalwnumber}
k_{1}^c=arcos(-\frac{U}{2V})  \;.
\end{equation}
An antiresonance appears only if the following inequalities are
satisfied: $\pi-k_1^c<\tilde k_1<k_1^c$ (here for simplicity we assume
both interaction constants positive). As it is seen from Fig.
\ref{gcorr4} the perturbative limit (\ref{111}) works well even in
case of presence of an antiresonance. Equation
(\ref{eq:resonance}) further tells that an antiresonance will be observed
even for $U=0$. In this case the seed
wavenumber $\tilde k_1=\pi/2$ is not modified by interaction $V$.
On the other hand if
$V=0$ the antiresonances are not observable.

For larger values of interaction constants the
perturbative predictions will get significant corrections.
To show this we plot three
dimensional graphs of weight function versus seed $\tilde k_1$ and
probe $k_1$ wave numbers for various values of interaction
constants in Fig. \ref{gcorr12}. As seen
for small values of interaction constants the track of the antiresonances
keeps the symmetry in the seed-probe mode number space traces predicted
by perturbation theory. However for large interaction constants this symmetry
is lost.

{\it Antisymmetric
states.} The structure of the antisymmetric matrix (\ref{eq:matasym})
suggests that the weight function for antisymmetric states can be
computed as:
\begin{equation}\label{form_asymweight}
C^a(k_1;\tilde{k}_1)
=\frac{V^2|\langle\Psi_{k_1}^0|\Psi_{\tilde{k}_1}^0\rangle|^2}
{(E_{\tilde{k}_1}^0-E_{{k}_1}^0)^2}.
\end{equation}
and according to this formula the weight function does not develop antiresonances.
This has been confirmed by numerical
diagonalization.

{\it Discussions.} Let us discuss the meaning of the observed
antiresonances. Two particles, when travelling along the lattice,
will meet, interact, and scatter. If prepared in an initial
symmetric noninteracting seed state, the particles will scatter
into all other available noninteracting symmetric states - except
for one special. This is because the scattering can go either via
the onsite interaction $U$ or via the intersite interaction $V$. A
corresponding destructive interference makes the amplitude in this
particular scattering state exactly zero. Antisymmetric states
have strict zero occupation on the same site, and therefore only
one scattering path (using $V$) is left. Consequently they do not
show antiresonances. But they will, if we add even more distant
(e.g. next-to-nearest-neighbor) interactions.

{\it Acknowledgments} J-P.  Nguenang and R. Khomeriki acknowledge
the warm hospitality of the Max Planck Institute for the Physics
of Complex Systems in Dresden.

\begin{widetext}.
\begin{figure}[t]
\includegraphics[width=6in]{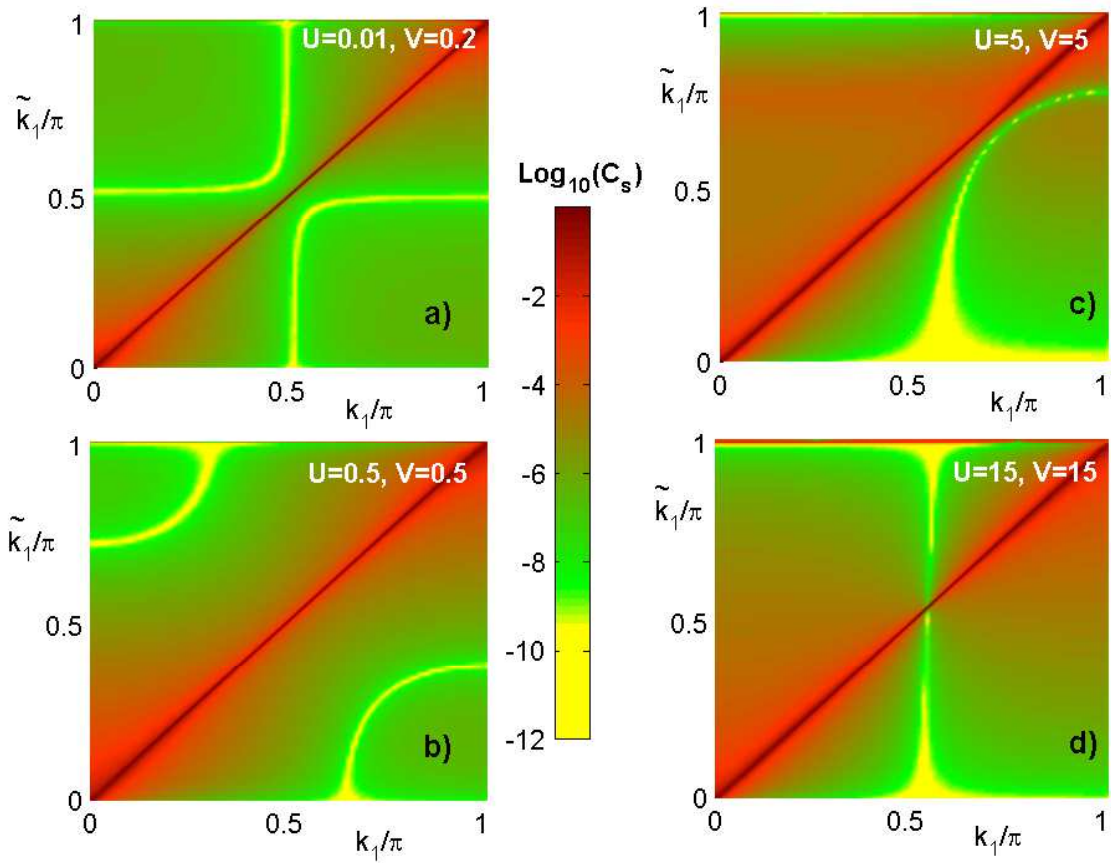}
\caption{\label{gcorr12} Three dimensional plots of the weight
function for symmetric states for a fixed value of the Bloch
wavenumber $k=0.12\pi$ and different interaction constants $U$ and
$V$. The lattice size is the same $f=101$ as in the previous
plots.}
\end{figure}
\end{widetext}

\end{document}